\documentclass[aps,pra, 10pt,twocolumn, superscriptaddress]{revtex4-2}
\usepackage{amsmath,amsfonts,amssymb}
\usepackage{graphicx}
\usepackage{bm}      
\usepackage[colorlinks=true,linktocpage=true,breaklinks=true,allcolors=blue]{hyperref}
\usepackage{xcolor}
\usepackage{ulem}

\begin{document}

\title{Chirality-induced spin selectivity due to an emergent molecular orbital moment}

\author{Sumit Ghosh}
\email{ghoshs@fzu.cz}
\affiliation{The Institute of Physics of the Czech Academy of Sciences, 162 00 Prague, Czech Republic}

\author{Angela Wittmann}
\email{a.wittmann@uni-mainz.de}
\affiliation{Institute of Physics, Johannes Gutenberg-University, 55128 Mainz, Germany}

\author{Frank Matthes}
\affiliation{Peter Gr{\"u}nberg Institut (PGI-6), Forschungszentrum J{\"u}lich GmbH, 52428 J{\"u}lich, Germany}

\author{Daniel E. B{\"u}rgler}
\email{d.buergler@fz-juelich.de}
\affiliation{Peter Gr{\"u}nberg Institut (PGI-6), Forschungszentrum J{\"u}lich GmbH, 52428 J{\"u}lich, Germany}

\begin{abstract}
The microscopic origin of chirality-induced spin selectivity (CISS), a phenomena where chiral molecules connected to a magnetic electrode show large spin-dependent conductance asymmetry even in ambient temperature, still remains an open question. Most of the existent theories are based on some sort of effective spin-orbit coupling or interfacial scattering, which do not provide satisfactory explanations for experimental observations such as the magnitude of spin polarization or the length dependence of the effect. We propose an alternative theoretical framework rooted in an intrinsic molecular contribution that shows a robust CISS effect and scales with the length of the molecule. The mechanism is based on the emergent molecular magnetic moment (MOM), which can produce experimentally observed point-symmetric \textit{I-V} curves without violating the Onsager-Casimir reciprocity relation. The MOM has an intrinsic component that is governed by the chirality of the molecule and leads to finite CISS magnetoresistance and magnetochiral conductance asymmetry (MChA) that are even in bias voltage. By decomposing the MOM into its Fermi sea and Fermi surface components, we show that under strong bias, the MOM can be dominated by the Fermi surface contribution, which causes a sign change in the MChA leading to a behavior similar to that of electric magnetochiral anisotropy (eMChA). Finally, we analyze the effect of a magnetic field within the same framework, which may provide better insight into the crossover between CISS and eMChA.
\end{abstract}

\maketitle

\section{Introduction}

Certain organic molecules respond differently to electrons with opposite spin which can be captured in transport or photoemission  experiments. Since the effect stems from their chirality, the phenomenon as well as the underlying physical mechanism are called \textit{chirality-induced spin selectivity} (CISS) \cite{Naaman2012, Naaman2019, Evers2022, Aragones2022, Foo2025}. Transport experiments on CISS are typically performed by contacting a chiral molecule with two electrodes, of which at least one is ferromagnetic. The current-voltage (\textit{I-V}) curves of such two-terminal devices are then compared either for opposite chirality of the system or for opposite magnetization directions of the ferromagnetic electrode. In both cases, the \textit{I-V} curves are split, and their asymmetries (difference divided by the sum) are the measures of the CISS effect and referred to as  magnetochiral conductance asymmetry (MChA = $\frac{I_+ - I_-}{I_+ + I_-}$, +/- denote opposite chirality of the molecule) and CISS magnetoresistance (CISS-MR = $\frac{I_\uparrow - I_\downarrow}{I_\uparrow + I_\downarrow}$, $\uparrow$/$\downarrow$ denote magnetization direction), respectively. Interestingly, the vast majority of such transport measurements report a crossing of the \textit{I-V} curves at zero bias, which means that MChA and CISS-MR are even functions of the bias. This is not only the case for numerous chiral molecules ({\it e.g.}, \cite{Xie2011, Kiran2016, Bloom2016, Aragones2016, Mishra2020, Safari2024, Nguyen2024, Aragones2025}), but also for chiral 2D perovskites \cite{Lu2020}, van der Waals superlattices intercalated with chiral molecules \cite{Bian2023}, nanofibers \cite{Kulkarni2020,Albro2025} and hybrid organic–inorganic thin films devices with embedded chirality \cite{Bustami2022}.

In general, CISS is explained as the result of atomic spin-orbit coupling (SOC) \cite{Gutierrez2012, Varela2016, Geyer2019, Dalum2019} or geometric SOC \cite{Vager2012, Shitade2020}. However, being one of the lightest elements, the intrinsic SOC in carbon is quite small even in the presence of a high curvature as in carbon nanotubes \cite{Kuemmeth2008}. Besides, these theories are based on continuum models which are applicable when the system length is significantly longer than the inter-atomic spacing and therefore is well suited for the CISS effect observed in DNA-like molecules \cite{Xie2011}. For small systems, due to a strong confinement effect, the continuum models may not properly describe the dynamics of the system. In addition, for a small radius of curvature, the one-dimensional relativistic theory shows divergent behavior \cite{Ghosh2013} which can alter the theoretical outcome significantly.

One of the major obstacles in the SOC-based theoretical explanations of CISS is posed by the reciprocity relations by Onsager-Casimir \cite{Onsager1931a, *Onsager1931b, *Casimir1945} and B\"uttiker \cite{Buttiker1988} which dictate that in absence of time reversal symmetry breaking interaction, the transport coefficients satisfy $\mathcal{G}_{ij}(\bm{B}) = \!\mathcal{G}_{ji}(\bm{-B})$. In presence of magnetization this can be further generalized to $\mathcal{G}_{ij}(\bm{B,M})=\mathcal{G}_{ji}(\bm{-B,-M})$ \cite{Shtrikman1965}. For a two-terminal device this implies that the longitudinal conductance $\mathcal{G}_{xx}$ is an even function of the magnetization resulting in zero magnetoresistance. Since SOC does not break time reversal symmetry, it can not directly give rise to a non-zero magnetoresistance. However, it can give rise to spin-dependent interfacial scattering or charge accumulation which can produce measurable CISS effects \cite{Yang2020}. A large number of theoretical models have been developed based on interactions at molecule-electrode interfaces \cite{Huisman2021, Liu2021, Alwan2021, *Monti2024, *Sarkar2025, Tirion2024, Zhao2025} as well as electron-phonon interaction and electron correlation \cite{Fransson2019, *Fransson2020, *Fransson2021}, which allow to circumvent the restrictions of the reciprocity relations. 
Although these models can explain numerous experimental observations, there are also several discrepancies \cite{Tirion2025, Fransson2026}. Spin filter-based theories yield non-symmetric and non-crossing \textit{I-V} curves which is in sharp contrast with the experimentally observed largely point-symmetric and crossing \textit{I-V} curves. Models based on interfacial scattering rely on the presence of molecule-electrode interactions which cannot explain the length dependence of the CISS effect \cite{Mishra2020}. Recent measurements using scanning tunneling microscopy (STM), where the STM tip acts as a weakly coupled electrode, also show that the magnitude of CISS does not  depend solely on the molecule-electrode coupling \cite{Safari2024, Nguyen2024}. Moreover, CISS has been observed even in absence of direct coupling to electrodes \cite{Eckvahl2023}. These observations clearly indicate that the molecule itself also plays a significant role in CISS \cite{Aragones2025}.

In this paper, we overcome these disparities by presenting an intrinsic mechanism for CISS arising from the molecule itself, which does not explicitly rely on any interfacial mechanism or atomic SOC and scales with the length of the molecule. The proposed mechanism is based on the circulation of charge current in molecular rings generating a \textit{molecular orbital moment} (MOM). This is inherently different from the atomic orbital angular moment. A similar ring current and magnetic moment have been previously predicted in biased quantum circuits \cite{Cini2010} and C$_{60}$ molecular bridges \cite{Nakanishi2001}. In a C$_{60}$ molecule, such ring current can be several tens times larger than the source-drain current flowing through the molecular bridge. However, the current and moments being governed by the bias voltage would produce non-crossing \textit{I-V} curves. Contrary to that, the MOM reported here has a bias-independent intrinsic component leading to the experimentally observed point-symmetric and crossing \textit{I-V} curves. The chirality of the molecule is imprinted on the electron dynamics via a vector field, which is determined by the molecular distortion \cite{Sasaki2005, Vozmediano2010}. The resulting ring current has contributions from quantum states located both near and far from the Fermi level. Their relative contribution determines the sign of the MOM, which can lead to additional crossings of the \textit{I-V} curves, especially at large bias voltage. In presence of an external magnetic field, its vector potential can couple to the chirality-induced emergent vector field, which allows us to extend  our formalism to investigate the impact of a magnetic field on the CISS effect \cite{Singh2025}.

\section{Results and discussion}

\begin{figure}[ht!]
\centering
\includegraphics[width=0.48\textwidth]{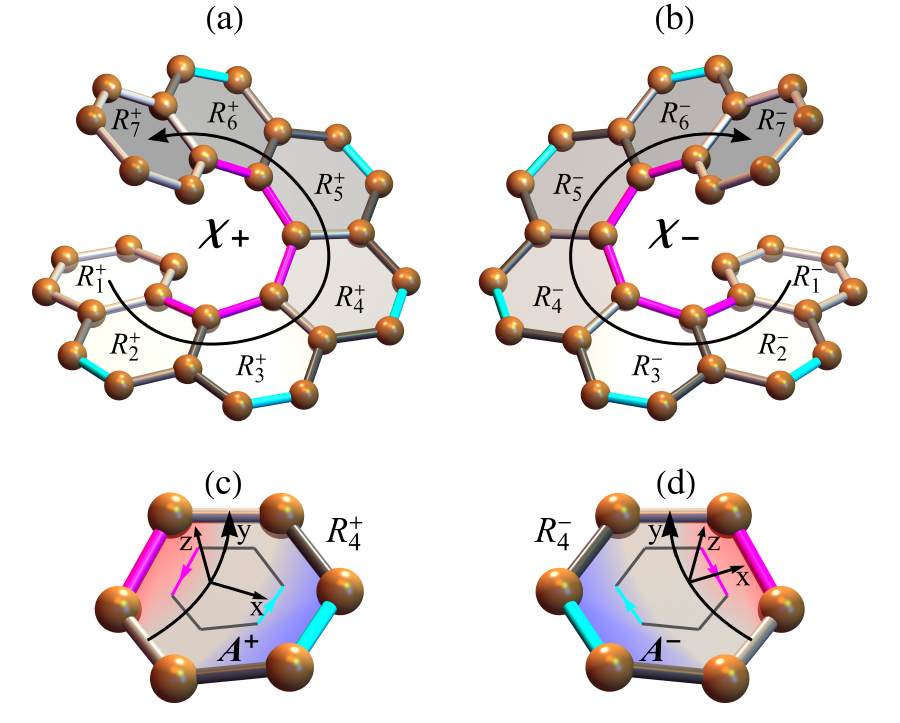}
\caption{Variation of the bond lengths and orientation of the vector field $\bm{A}^\chi$ in the [7]H enantiomers. (a) Right-chiral ($\chi_+$) and (b) left-chiral ($\chi_-$) [7]H enantiomers, where $R_1$-$R_7$ denote the constituent C six-rings. Cyan and magenta colors represent shorter and longer bonds and gray represents intermediate bonds. Solid black arrows indicate the direction of electron motion for positive current. (c) and (d) show the variation of bond length in the $4^{th}$ C six-ring ($R^\pm_4$), where $x$, $y$, and $z$ span the local Cartesian coordinate system in each ring. The gradient of the bond length is shown in color in the plane of the C six-ring. Cyan and magenta arrows show the in-plane component of $\bm{A}^\chi$ along the outer and inner edge as a representative case.}
\label{fig:h7}
\end{figure}

For our analysis, we focus on heptahelicene ([7]H) molecules as a prime helical candidate for CISS. The [7]H molecule is made of seven benzene rings connected along an arc and can occur in two enantiomers, namely right-handed ($\chi_+$) or left-handed ($\chi_-$, in the following we use a superscript $\chi=\pm$ to denote whether an entity belongs to the  $\chi_+$ or $\chi_-$ [7]H molecule). Figure\,\ref{fig:h7} shows the relaxed configurations of the C atoms in isolated $\chi_+$ and $\chi_-$ molecules obtained from first-principle calculations. In the case of a planar hexagonal structure, such as graphene, all C-C bonds have the same length (1.42\,\rm{\AA}). However for $\chi_\pm$ [7]H molecules, the bond lengths show a systematic variation (Figs.\,\ref{fig:h7}a,b, see Supplementary Note 1 for the exact structure with the H atoms and the different bond lengths). Independent of their handedness, the inner edge is made of longer (magenta) bonds and the outer edge is made of shorter (cyan) bonds \cite{Navaza1979, Murguly2000}. Such a variation of the bond lengths can be modeled as a vector field that can be incorporated as a Peierls phase \cite{deJuan2011} in the Hamiltonian of the system. In the continuum limit this field can be interpreted as a gauge field \cite{Vozmediano2010, Shapourian2015}. To understand the orientation of this field let us focus on the central C six-ring ($R^\chi_4$). Let us consider a common Cartesian coordinate system for both molecules such that the positive $y$ axis denotes the direction $R^\chi_4 \rightarrow R^\chi_5$, positive $z$ axis denotes the normal to the plane of the C six-ring and the $x$ axis is mutually perpendicular to both of them (see Fig.\,\ref{fig:h7}c,d). In this common reference frame, the bond length gradient (red to blue regions in Fig.\,\ref{fig:h7}c,d) is along $\mp x$ for the $\chi_\pm$ [7]H molecule. Since the corresponding vector field ($\bm{A}^\pm$) is proportional to this gradient \cite{Sasaki2005, Vozmediano2010,deJuan2011, Shapourian2015}, the loop integral $\oint_\chi\!\bm{A}^\chi\!\cdot\!d\bm{l}$, where the closed integration path runs in counter-clockwise direction along the C six-ring, in $\chi_+$ will be opposite to the that in $\chi_-$, which can lead to distinct transport features in $\chi_\pm$ molecules. Here we choose the vector field in such a way that $\oint_\chi\!\bm{A}^\chi\!\cdot\!d\bm{l}\!=\!\chi 0.06 \frac{h}{e}$. The magnitude is estimated from the variation of the bond length $(l_{\rm max}-l_{\rm min})/l_{\rm avg}\!=\!0.06$, where $l_{\rm min}\!=\!1.366$\,{\AA}, $l_{\rm max}\!=\!1.454$\,{\AA} and $l_{\rm avg}\!=\!1.415$\,{\AA} according to our ab-initio calculation (see Supplementary Note 1 for details).

\subsection{Theoretical model of [7]H molecule}

We can now map the [7]H molecules onto a tight-binding model considering one orbital per site as
\begin{eqnarray}
H^\chi  = \sum_i \epsilon^\chi_i c_i^\dagger c_i + t \sum_{\langle ij \rangle} e^{i\frac{e}{\hbar}\bm{A}^\chi_{ij} \cdot \bm{r}_{ij}} c_i^\dagger c_j + H.C.,
\label{H}
\end{eqnarray}
where $\epsilon^\chi_i$ is the onsite energy at site $i$. For brevity, we ignore the H atoms and only consider C-C hopping defined by the hopping parameter $t$. The onsite energies ($\epsilon^\chi_i$) are initially set to zero. $\bm{r}_{ij}$ is the spatial vector from atom $i$ to atom $j$. Without any loss of generality, we divide all bonds into three classes - (i)\,inner (magneta bonds in Fig.\,\ref{fig:h7}), (ii)\,outer (cyan bonds in Fig.\,\ref{fig:h7}) and (iii)\,intermediate (grey bonds in Fig.\,\ref{fig:h7}). Considering the bond length variations obtained from the first-principle calculation, which is incorporated through $\bm{A}^\chi$, we assume that $\bm{A}^\chi$ is finite in $R^\chi_2$ to $R^\chi_6$ and zero in $R^\chi_1$ and $R^\chi_7$. For $R^\chi_2$ to $R^\chi_6$, we choose $\bm{A}^\chi$ to be finite along the cyan and magenta bonds (see Fig.\,\ref{fig:h7}c,d) and zero otherwise. This choice is not unique and any other choice that produces the same loop integral ($\oint_\chi\!\bm{A}^\chi\!\cdot\!d\bm{l}$) for a C six-ring is equally valid.

To simulate the device configuration for STM/AFM measurements, the molecule is attached to suitable bottom and top electrodes. This device configuration is used to calculate its transport properties (see Methods section for details). The calculations are carried out at approximately 58\,K (5\,meV) implemented by the Fermi-Dirac distribution. This temperature is chosen to ensure better convergence of the energy integration and does not have any other physical consequence in our calculation. A similar result is obtained for room temperature, at which most CISS transport experiments are performed \cite{Xie2011, Nguyen2024, Aragones2025}, and also for the measurement temperature of 5\,K in \cite{Safari2024}.

\subsection{Ring current, MOM and Onsager-Casimir reciprocity}

The molecular structure allows the possibility of \textit{ring currents} which can flow inside the hexagonal C six-rings. Such currents do not contribute to the net current coming out of the device and therefore can not be measured directly in a transport experiment.  
In presence of finite bias the \textit{bond current} flowing between two atoms can be calculated using the non-equilibrium Green's function ($G^<$, see Methods section for details), and the effective \textit{molecular ring current} can be obtained by summing over all bonds present in a ring along the counter-clockwise direction (see Methods section for details).  The effective magnetic dipole moment or the MOM due to this ring current is given by the product of the ring current and the area of the hexagon. For evaluating the area, we assume that locally each six-ring is in the $xy$ plane defined by a regular hexagon with a side length of 1.42\,{\AA}, and that the $z$ axis represents the normal to the plane. Therefore the induced MOM will be aligned along the $\pm \hat{z}$ direction in its local frame (see Fig.\,\ref{fig:h7}c,d). This MOM is assumed to be equally distributed among all six atoms in the ring. The interaction of such localized magnetic moments with the spin magnetic moment of a conduction electron can be incorporated in the Hamiltonian (Eq.\,\ref{H}) as a modification of the onsite energy with a double exchange-type interaction \cite{Anderson1955} as
\begin{eqnarray}
\epsilon^\chi_i \to \epsilon^\chi_i - J_M (\bm{M}^\chi_i \cdot \bm{s})
\label{Ms}
\end{eqnarray}
where $J_M$ is the effective Hund coupling, $\bm{M}^\chi_i$ is the current-induced MOM at site $i$ and $\bm{s}$ is the spin magnetic moment of the electron. Unless explicitly mentioned, here we always consider $\bm{s}$ to be along +$\hat{z}$ direction. The change in $\epsilon^\chi_i$ will also cause a change in the current and consequently in the MOM. Therefore these two quantities are calculated self-consistently until both of them attain the desired level of convergence.

This interaction does not arise from any relativistic effect. Therefore its impact can be significantly large compared to the effect produced by the atomic SOC. Here we choose $J_M\!=\!+1$\,eV/$\mu_B^2$ which is estimated from the intra-atomic exchange integral \cite{Anderson1955}. The positive sign of $J_M$ indicates that the spin of the electron and the MOM prefer to be parallel. Depending on the other microscopic details, $J_M$ can be negative in certain materials. However, choosing a negative $J_M$ only swaps the $I$-$V$ curves of $\chi_+$ and $\chi_-$ molecules, leaving their qualitative behavior unchanged. Note that Eq.\,\ref{Ms} predicts that the degeneracy of transport properties can be lifted in two ways - (i) with different $\bm{M}^\chi$ while keeping $\bm{s}$ fixed (finite MChA) or (ii) with different $\bm{s}$ while keeping $\bm{M}^\chi$ fixed (finite CISS-MR) and therefore can explain these two scenarios within the same framework. In the following, we will explain all features of the MChA, from which the properties of the CISS-MR can also be derived.

\textbf{Onsager-Casimir reciprocity:} Note that the finite MChA and CISS-MR do not violate Onsager-Casimir reciprocity within this framework. The interaction $\bm{M}^\chi_i \cdot \bm{s}$ preserves the time-reversal symmetry leading to $\mathcal{G}_{xx}(\bm{M}^\chi ,\bm{s})\!=\!\mathcal{G}_{xx}(\bm{M}^\chi \cdot \bm{s})\!=\!\mathcal{G}_{xx}(-\bm{M}^\chi \cdot -\bm{s})\!=\!\mathcal{G}_{xx}(-\bm{M}^\chi ,-\bm{s})$. Consequently the conductance can have an odd contribution from both $\bm{M}^\chi$ and $\bm{s}$ resulting in a finite MChA and CISS-MR, respectively.
A similar interaction, in which the charge current is coupled to an external magnetic field, has been conjectured by Rikken \textit{et al.} \cite{Rikken2001}. It produces non-crossing \textit{I-V} curves  and was termed {\it electric magnetochiral anisotropy} (eMChA). Such an interaction would cause a sign change of the MChA at zero bias. A similar scenario can also be observed within our framework in presence of large bias voltages. However, for such a large bias voltage, the system is far from equilibrium and Onsager-Casimir reciprocity relations no longer hold.

\subsection{Symmetric MOM and crossing {\it I-V} curves}

\begin{figure}[ht!]
\centering
\includegraphics[width=0.48\textwidth]{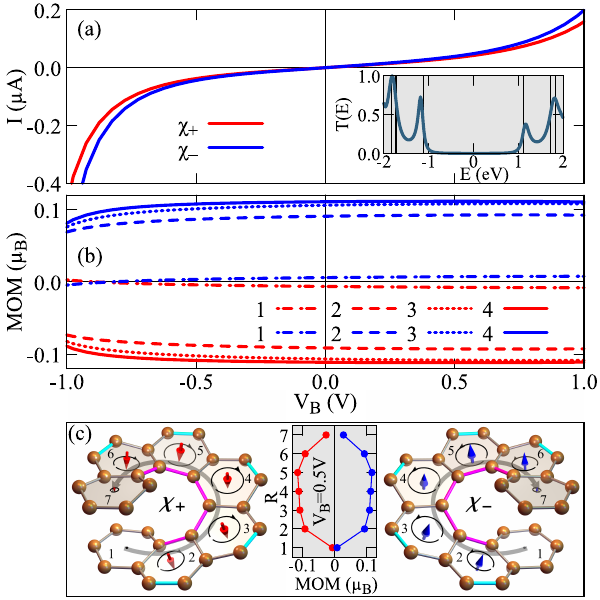}
\caption{\textit{I-V} characteristics for electrons with +$\hat{z}$ spin-polarization and corresponding MOMs. The Fermi level ($E_F$) is at 0.5\,eV. (a) \textit{I-V} curves of $\chi_+$ (red) and $\chi_-$ (blue) molecules (Fig.\,\ref{fig:h7}a,b). Inset shows their degenerate transmission coefficient where the vertical black lines denote the eigenvalues of isolated molecule. (b) MOM in the molecular rings $R^{\chi}_1$ to $R^{\chi}_4$ for $\chi_+$ (red) and $\chi_-$ (blue) molecules as a function of the bias voltage. (c) Schematic of MOMs and net current in $\chi_+$ (left) and $\chi_-$ (right) molecules at $V_B=0.5$\,V, where red ($\chi_+$) and blue ($\chi_-$) arrows denote the MOM and thick gray arrow show the direction of net current. Inset shows the magnitudes of the MOMs.}
\label{fig:IV}
\end{figure}

For STM/AFM measurements we are interested in the transport properties of the molecule within the bias window, which typically spans $\pm 0.5$\,eV around $E_F$.  To begin with, the effective Fermi level of the device is set to $E_F=0.5$\,eV taking into account charge transfer from the electrode to the molecule. The \textit{I-V} curves if the two enantiomers (Fig.\,\ref{fig:IV}a) show a crossing at zero bias, which is a salient feature of numerous experimental results \cite{Xie2011, Kiran2016, Bloom2016, Aragones2016, Mishra2020, Safari2024, Nguyen2024, Lu2020, Bian2023, Kulkarni2020, Bustami2022, Aragones2025,Albro2025}. The transmission coefficient or the zero-bias conductance remains degenerate for both enantiomers as expected from Onsager-Casimir reciprocity relations (see inset of Fig.\,\ref{fig:IV}a). The net current and MOMs are calculated self consistently (see Method section for details). Note that, although the magnitude of current shown in our results has been observed in few cases \cite{Singh2025}, it is significantly higher than the typical values in STM/AFM experiments \cite{Xie2011,  Kiran2016, Safari2024, Nguyen2024}. This is due to strong molecule-electrode couplings used here for a better understanding of the underlying physical mechanism. By reducing these couplings each by a factor of five one can reduce the net current by three orders of magnitude and reach the typical experimental level (see Supplementary Note 2).

Figure \ref{fig:IV} shows that although the net current through the molecule varies significantly and even changes sign with the applied bias, the MOM remains largely unaffected by the bias, and points in opposite directions for $\chi_+$ and $\chi_-$ molecules. These features reflect its intrinsic nature (see Supplementary Note 3 for a qualitative discussion). Note that we have set $\bm{A}^\chi=0$ for $R^\chi_1$ and $R^\chi_7$. The  MOM in these rings results from the proximity effect (Fig.\,\ref{fig:IV}c), which is why their magnitude is considerably smaller (see Fig.\,\ref{fig:IV}b). These MOMs are coupled to the spin of the conduction electron via Eq.\,\ref{Ms} which lifts the degeneracy of the \textit{I-V} curves for $\chi_+$ and $\chi_-$ molecules, resulting in a finite MChA. Note that, near equilibrium ($V_B \rightarrow 0$), $\bm{M}^+_i = -\bm{M}^-_i$, and therefore replacing $\chi_+$ [7]H with $\chi_-$ [7]H will have the same effect as the reversing the MOM ($\bm{M}^\chi_i \rightarrow -\bm{M}^\chi_i$). Consequently, the order of \textit{I-V} curves will reverse, resulting in a sign change of the MChA. The order of the \textit{I-V} curves is also reversed when the spin of the injected electron is flipped ($\bm{s} \rightarrow -\bm{s}$) by reversing the magnetization of the electrode. This gives rise to a finite CISS-MR, whose sign is determined by the chirality of the molecule. In fact, all these features have been observed experimentally  \cite{Aragones2016, Safari2024}.

\begin{figure}[ht!]
\centering
\includegraphics[width=0.48\textwidth]{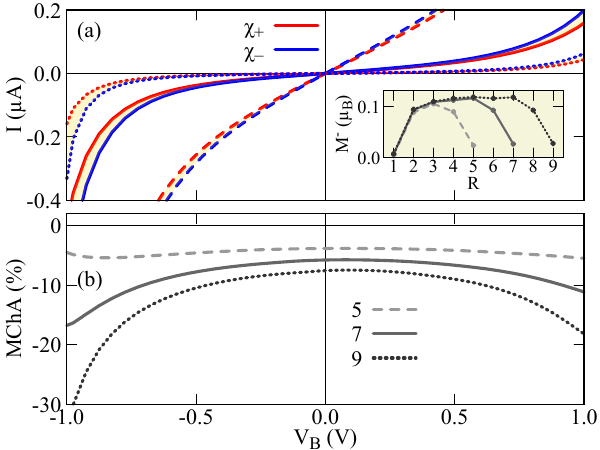}
\caption{Length dependence of the CISS effect. (a) \textit{I-V} curves for [5]H (dashed), [7]H (solid) and [9]H (dotted) molecules, where red and blue represent $\chi_+$ and $\chi_-$ molecules, respectively. Inset shows the MOMs for the $\chi_-$ enantiomers at $V_B=0.5$\,V.  The Fermi level for [7]H is at 0.5\,eV, and for [5]H and [9]H it is set at an equal distance from the lowest unoccupied state. (b) shows the MChA for these three molecules.}
\label{fig:IL}
\end{figure}

\textbf{Length dependence:} Since each molecular ring contributes to lifting the degeneracy of the \textit{I-V} curves, it is natural to ask whether the effect scales with the number of rings. To investigate the impact of the length, we consider three helicene molecules with 5, 7 and 9 rings and evaluate the corresponding \textit{I-V} curves and MOMs (Fig.\,\ref{fig:IL}). As can be seen from Fig.\,\ref{fig:IL}a, the net current through the molecule decreases as its length increases, as expected for transmission through tunneling barrier. Concomitantly, the net MOM also grows with increasing molecular length, which increases the total resistance for a given spin channel, resulting in an increase of the MChA (Fig.\,\ref{fig:IL}b). The CISS effect thus scales within our theoretical framework with the length of the molecule, which has been observed experimentally as well \cite{Mishra2020}.

\subsection{Fermi surface and Fermi sea contributions to MOM and finite-bias crossing of \textit{I-V} curves}

\begin{figure}[h!]
\centering
\includegraphics[width=0.48\textwidth]{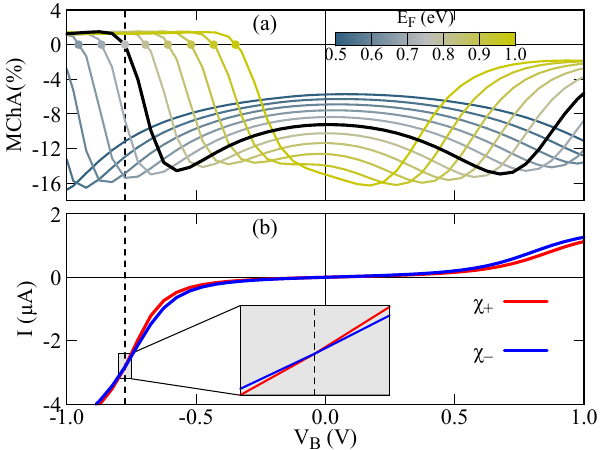}
\caption{MChA for different Fermi energies and occurrence of second crossings. (a) MChA for different Fermi energies marked by different colors (see inset for the color code). The thick black line shows the MChA for $E_F=$0.75\,eV, and the vertical dashed line marks the voltage at which the MChA changes sign. (b) \textit{I-V} curves for $E_F=$0.75\,eV, where the red and blue lines represent $\chi_+$ and $\chi_-$ enantiomers,respectively. Inset shows the enlarged view of the crossing of the \textit{I-V} curves at finite bias.}
\label{fig:IVF}
\end{figure}

Additional contributions to the current coming from states near the Fermi level can occur when the Fermi level is closer to a resonance level, which can cause significant changes in the \textit{I-V} characteristics. 
To understand this effect, we shift the Fermi level gradually towards the resonant level at 1\,eV (see Fig.\,\ref{fig:IV}a, inset) and quantify the \textit{I-V} characteristics using the MChA (Fig.\,\ref{fig:IVF}).

As can be seen from Fig.\,\ref{fig:IVF}a, depending on the Fermi level, the MChA changes sign at a certain bias voltage. The closer the Fermi level is to the resonant level, which is at +1\,eV in this case, the smaller the bias voltage required to flip the MChA sign. In the \textit{I-V} curves, this is manifested as a second crossing (Fig.\,\ref{fig:IVF}b). 

\begin{figure}[ht!]
\centering
\includegraphics[width=0.48\textwidth]{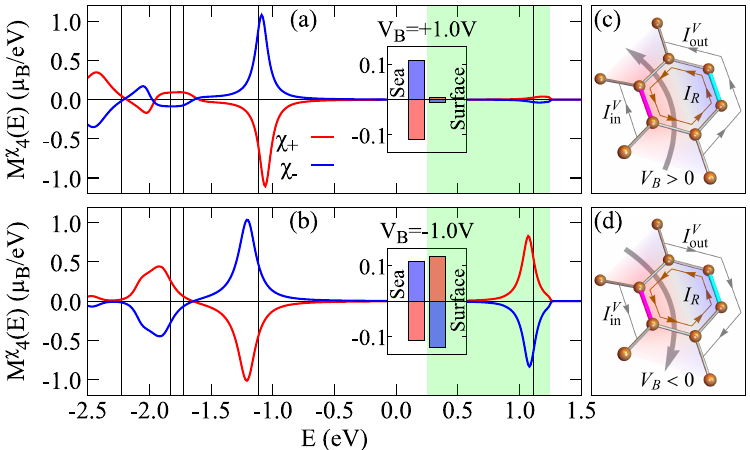}
\caption{Fermi sea and surface contributions to the MOM in $4^{th}$ ring for $E_F=0.75$\,eV. (a) and (b) show the energy-resolved contributions to the MOM for $\chi_+$ (red) and $\chi_-$ (blue) molecules at $V_B=+1$\,V and $V_B=-1$\,V respectively. The green regions denote the bias window and the vertical black lines the energy eigenvalues of the isolated molecule. Insets show the total Fermi sea and surface contributions. (c) and (d) show  schemes of the ring current ($I_R$) and the bias-dependent current ($I$, gray arrow) flowing in a six-ring of a $\chi_+$ molecule for positive and negative bias, respectively. $I$ flows partly along the inner ($I^V_{in}$) and outer ($I^V_{out}$) edges of the helix.} 
\label{fig:VS}
\end{figure}

Let us consider, as a representative case, the MOM of the $4^{th}$ six-ring ($\bm{M}^\chi_4$) with $E_F=0.75$\,eV  and $V_B=\pm1$\,V  to gain clear physical insight of the cause of this sign flip. To understand the variation of the MOM with  bias voltage, we decompose the MOM into two main components - (i) contribution coming from the bias window ($E_F-|V_B|/2 \leq E \leq E_F+|V_B|/2$) and (ii) the remaining contribution coming from rest of the occupied energy levels ($-\infty \leq E \leq E_F-|V_B|/2$). For the limiting case $V_B \to 0$, the former is known as the Fermi surface contribution, whereas the latter is known as the Fermi sea contribution \cite{Ghosh2019}. For compatibility, we are going to use the same terminology for finite bias voltage as well. The contribution from the Fermi sea is immune to external bias and accounts for the robust intrinsic contribution to the MOM (see Fig.\,\ref{fig:VS}). If the bias window is far from the resonant levels, the MOM is completely dominated by the Fermi sea resulting in a bias-independent MOM (see Fig.\,\ref{fig:IV}b). If the Fermi window comprises states in the vicinity of resonant levels, the contribution from the Fermi sea can be quite large (see Fig.\,\ref{fig:VS}b). Depending on the parity of different quantum states, the Fermi surface contribution can be opposite to the Fermi sea contribution \cite{Ghosh2019} resulting in an overall decrease in the MOM (see Fig.\,\ref{fig:CM}). 

When the Fermi surface contribution completely compensates for the Fermi sea contribution, the total MOM vanishes (Figs.\,\ref{fig:CM}c,e), and the same net current results for both $\chi_+$ and $\chi_-$ molecules. A further increase of the Fermi sea contribution causes a sign change of the MChA and a second crossing of the {\it I-V} curves (see Fig.\,\ref{fig:IVF}), which has also been observed experimentally \cite{Nguyen2024, Singh2025}. Near a resonant level, the Fermi surface contribution increases significantly, so that the second crossing can be achieved at smaller bias voltage.  In fact, if we set $\bm{A}^\chi=0$, the MOM is entirely contributed by the Fermi sea, and the second crossing shifts to zero bias, resulting in non-crossing \textit{I-V} curves (see Supplementary Note 4 for details). Therefore, the relative magnitude of Fermi sea and Fermi surface contributions can be exploited to reverse the sign of the MChA, thereby establishing a better connection between CISS and eMChA.

\subsection{Antisymmetric component of MOM}

\begin{figure}[ht!]
\centering
\includegraphics[width=0.48\textwidth]{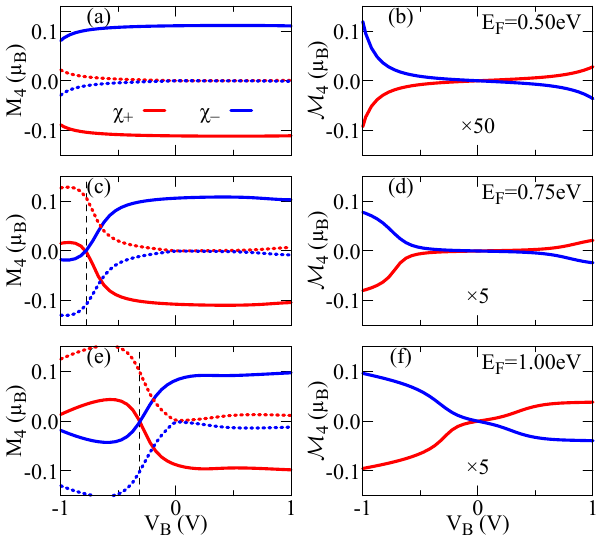}
\caption{Variation of the total MOM and its antisymmetric component. Total MOM (solid) and the Fermi surface contribution (dotted) of the $4^{th}$ six-ring ($\bm{M}^\chi_4$) for $E_F$ at (a)\,0.5\,eV, (c)\,0.75\,eV and (e)\,1.0\,eV, where red and blue lines represent the $\chi_+$ and $\chi_-$ enantiomers. The vertical dashed lines in (c,e) indicate the location where $\bm{M}^\chi_4=0$ and MChA vanishes (see Fig.\,\ref{fig:IVF}). (b), (d) and (f) show the corresponding antisymmetric component of the MOM (see Eq.\,\ref{MV}).
}
\label{fig:CM}
\end{figure}

For a better understanding of the bias-dependent component of the MOM, we consider three cases with Fermi energies 0.5\,eV, 0.75\,eV and 1.0\,eV and focus again on $M^\chi_4$ (Fig.\,\ref{fig:CM}). As discussed earlier, if the Fermi level is far from resonant levels, the MOMs are governed by the Fermi sea contribution (see Fig.\,\ref{fig:CM}a). Near a resonant level, the Fermi surface contribution becomes stronger (Figs.\,\ref{fig:CM}c,e). Note that the Fermi surface contribution of the MOM (dotted in Figs.\,\ref{fig:CM}a,c,e) does not change sign with the bias voltage, even though a bias-dependent current $I=I^{V}_{out}+I^{V}_{in}$ is also expected (see schemes in Figs.\,\ref{fig:VS}c,d). This is due to the influence of $\bm{A}^\chi$. To exclude its effect and extract the bias-dependent component we proceed in following way. Considering that the intrinsic component of the MOM is governed by $\bm{A}^\chi$, which changes sign with the handedness of the molecule, the intrinsic component must follow the equality $\bm{M}^{+}_i(\bm{A}^{+})$ = $\bm{M}^{-}_i(-\bm{A}^{-})$. Based on that, the bias-dependent antisymmetric contribution $\bm{\mathcal{M}}^\chi_4$ can be defined as
\begin{eqnarray}
\bm{\mathcal{M}}^{\pm}_i = \bm{M}^{\pm}_i(\bm{A}^{\pm}) - \bm{M}^{\mp}_i(-\bm{A}^{\mp}).
\label{MV}
\end{eqnarray}

As can be seen from Figs.\,\ref{fig:CM}b,d,f, $\bm{\mathcal{M}}^\chi$ is  antisymmetric in nature and one or two orders of magnitude smaller compared to $\bm{M}^\chi$. Note that this antisymmetric contribution is proportional to both the bias voltage -- and thus net current $I$ -- and the handedness $\chi$ of the molecule, $\bm{\mathcal{M}}^\chi \propto \chi I \bm{e}_z$ (Fig.\,\ref{fig:CM}b,d,f). Thus, the corresponding interaction term (Eq.\,\ref{Ms}) shows the proportionality $\bm{\mathcal{M}}^\chi \cdot \bm{s} \propto \chi I s$, similar to the conjecture of Rikken \textit{et al.} for eMChA \cite{Rikken2001}, and likewise can lead to non-crossing \textit{I-V} curves. It is worth mentioning that although small in magnitude,  $\bm{\mathcal{M}}^\chi$ can be present even in absence of any emergent vector field $\bm{A}^\chi$ and therefore can play a governing role in spin selectivity in a large class of molecules (see Supplementary Note 4 for details).

\subsection{Effect of external magnetic field}

\begin{figure}[h!]
\centering
\includegraphics[width=0.48\textwidth]{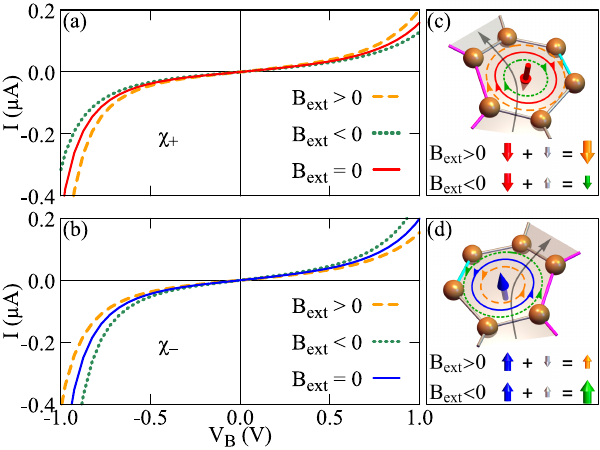}
\caption{\textit{I-V} curves under opposite external magnetic field for $\chi_\pm$ molecules. Here we use $|\oint\!\bm{A}^B\cdot d\bm{l}|=0.1 |\oint\!\bm{A}^\chi\cdot d\bm{l}|$. (a) and (b) show the $I$-$V$ curves for $\chi_+$ and $\chi_-$ enantiomers, where orange dashed and green dotted lines correspond to positive ($+\bm{B}_{ext}$) and negative ($-\bm{B}_{ext}$) external magnetic field, and red and blue lines show the \textit{I-V} curves without $\bm{B}_{ext}$ (see Fig.\,\ref{fig:IV}a). (c) and (d) show the schematic diagrams of the resulting ring current in the $4^{th}$ six-ring for positive bias voltage using the color code in (a). Bottom panels of (c,d) show the formation of the total MOMs, where the gray arrows show the $\bm{B}_{ext}$-induced MOMs and the red/blue arrows the ring current-induced MOMs.}
\label{fig:IVB}
\end{figure}

Finally, we explore the situation in presence of a magnetic field. For the time being we keep aside the complexity arising from the anisotropic magnetic susceptibility of aromatic molecules \cite{Musher1965} and consider a simple scenario in which an external magnetic field ($\bm{B}_{ext}$) is applied along the $z$ axis (see Fig.\,\ref{fig:h7}c,d) which can be incorporated as an additional Peierl's phase $e^{i \frac{e}{\hbar} \bm{A}^B \cdot \bm{r}_{ij}}$ to the hopping term in Eq.\,\ref{H}, where $\bm{A}^B$ is the vector potential for $\bm{B}_{ext}$. Note that while the integral $\oint_\chi\!\bm{A}^\chi\cdot d\bm{l}$ changes sign for molecules with opposite chirality ($\oint_\chi\!\bm{A}^\chi\cdot d\bm{l} \propto\chi$, see also Fig.\,\ref{fig:h7}c,d), the integral $\oint_\chi\!\bm{A}^B\cdot d\bm{l}$ is independent of the chirality and only changes sign with the direction of field ($\oint_\chi\!\bm{A}^B\cdot d\bm{l}\propto\bm{B}_{ext}\cdot\hat{\bm e}_z$). Similar behavior can be observed in the resulting ring current and MOM as well ( Fig.\,\ref{fig:IVB}). For $|\oint\!\bm{A}^B\cdot d\bm{l}| < |\oint\!\bm{A}^\chi\cdot d\bm{l}|$, the sign of the MOM will still be governed by $\bm{A}^\chi$, while the splitting of the  \textit{I-V} curves is produced by the $\bm{B}_{ext}$-induced MOM (see Figs.\,\ref{fig:IVB}c,d). Therefore the order of the \textit{I-V} curves will be reversed for molecules with opposite chirality (see Figs.\,\ref{fig:IVB}a,b) \cite{Bustami2022}. Furthermore, depending on the Fermi energy and bias voltage, the contributions from the Fermi sea and surface can also alter the sign of the MOM, resulting in secondary crossings (see Fig.\,\ref{fig:IVF}b) also in presence of an external magnetic field, which has been observed in recent experiments \cite{Singh2025}.

\section{Conclusion}

In this paper, we present an intrinsic mechanism for CISS in helical molecules. To demonstrate this we consider the two enantiomers of heptahelicene molecules and analyzed their transport properties. Our proposed mechanism for CISS relies on the emergent current-induced MOM. The chirality of the molecule is incorporated via a vector field obtained from the distortion of the molecule, which results in pairs of spin- or chirality-split point-symmetric \textit{I-V} curves. We further decompose the net MOM into a Fermi sea and Fermi surface contribution and show how the external bias voltage can alter the Fermi surface contribution and manipulate the MOM. The competition between the Fermi sea and Fermi surface contributions to the MOM gives rise to additional crossings of the \textit{I-V} curves at finite bias. The position of this secondary crossing can be adjusted through the proximity of the bias window to the resonant levels.  We also demonstrate the existence of an antisymmetric part of the MOM, which can be dominating in presence of a large bias current. We show that the coupling between the antisymmetric component of the MOM and the injected spin has a form similar to that proposed by Rikken \textit{et al.} \cite{Rikken2001} in the context of electric magnetochiral anisotropy (eMChA). Finally, we show the effect of an external magnetic field leading to a magnetic field-dependent splitting of the \textit{I-V} curves, which can reverse their order for opposite chirality of the molecule. Within a reasonable range of realistic parameter values, our results qualitatively agree with a wide range of experimental findings. The formalism, although developed for a helical molecule, can be generalized for other more complex chiral molecules, which may reveal even more interesting aspects of the CISS effects.

\section{Methods}

\subsection{First-principle calculation}

The structure relaxations of the molecules have been done with Quantum Espresso \cite{Giannozzi2009} using SG15 pseudopotentials \cite{Hamann2013}. A cell of dimension 25$\times$25$\times$25 \AA$^3$ is used to ensure that the relaxations are free from the impact of the periodic boundary conditions. We use a cutoff of 50\,Ry for the wavefunctions and 200\,Ry for the charge density. The structures have been relaxed until they reach an energy convergence threshold of 10$^{-5}$\,a.u. and the forces on all atoms are less than 10$^{-4}$\,a.u.

\subsection{Device configuration and non-equilibrium transport}

The [7]H molecules are defined by a single-orbital tight-binding Hamiltonian
\begin{eqnarray}
H^\chi  = \sum_i \epsilon^\chi_i c_i^\dagger c_i + t \sum_{\langle ij \rangle} e^{i\frac{e}{\hbar}\bm{A}^\chi_{ij} \cdot \bm{r}_{ij}} c_i^\dagger c_j + H.C.,
\label{H1}
\end{eqnarray}
considering the C atoms only. $\epsilon^\chi_i$ is the onsite energy, $t$ is the C-C hopping parameter which we set to -2.7\,eV, considering the C-C hopping parameter in graphene. $\bm{A}^\chi_{ij}$ is the vector field due to structural distortion (see Fig.\,\ref{fig:h7}c,d for its orientation). $\bm{r}_{ij}$ is the vector connecting site $i$ with site $j$. $\chi$=$\pm$1 denote the right ($\chi_+$) and left ($\chi_-$) chiral [7]H molecule.

To simulate the device configuration, we attach the molecule to a wider (6$\times$6) bottom electrode and a narrow (2$\times$2) top electrode, which represents the conductive tip of a scanning tunneling microscope (STM) or an atomic force microscope (AFM). The electrodes are mapped onto a square lattice with zero onsite energy and a hopping parameter of $-1.5$\,eV which is suitable for metallic electrodes. Considering the distance between the molecule and the electrodes, the coupling between the molecule and the bottom electrode is chosen to be $-1.5$\,eV, while the coupling between the molecule and the top electrode is set to $-0.5$\,eV. In practice, the former coupling strength depends on the chemical bonding to the substrate and the latter decays exponentially with the distance between the molecule and the STM/AFM tip. 

We also assume that only the first C six-ring is connected to the bottom electrode and that the top electrode is weakly coupled to the two closest C atoms. The effects of these semi-infinite electrodes are incorporated via their retarded self-energy $\Sigma^R_{B,T}$, where the subscripts  $B$ or $T$ correspond to the bottom or top electrode, which is calculated using the recursive Green's function method \cite{Lewenkopf2013}. We use a constant Gaussian broadening of 5\,meV which is well suited for the metallic interfaces. The retarded Green's function for the molecule is given by $G^R(E)\!=\![E - H^\chi -(\Sigma^R_B + \Sigma^R_T)]^{-1}$. The transmission coefficient, which also represents the dimensionless conductance of the device at zero bias, is given by $T(E)\!=\!Tr[G^R \Gamma_B G^A \Gamma_T]$ \cite{Datta1995} where $G^A$=$(G^R)^\dagger$ is the advanced Green's function and $\Gamma_{B,T}\!=\!i(\Sigma^R_{B,T} - {\Sigma^R_{B,T}}^\dagger)$. 

In the presence of a finite bias $V_B$ we use an onsite energy $\pm eV_B/2$ for the bottom/top electrode and a linear potential gradient for the molecule. The resulting current through the device is given by $I\!=\!\frac{e}{h}\int_{-\infty}^{+\infty} T(E)(f_B(E)-f_T(E)) dE$ where $f_{B,T}(E)\!=\!1/(1+e^{(E -(E_F \pm eV_B/2))/(k_BT)})$ is the Fermi-Dirac distribution for the bottom/top electrode and $E_F$ the Fermi energy. Here we choose $k_BT\!=\!5$\,meV which is approximately 58\,K. 

The non-equilibrium properties of the molecule is obtained from the lesser Green's function of the molecule defined as $G^{<}\!=\!i G^R (f_B \Gamma_B$+$f_T \Gamma_T) G^A$.  The bond current from atom $i$ to atom $j$ is given by
\begin{eqnarray}
I_{ij} = \frac{e}{h} \int_{-\infty}^{+\infty} dE \left( H^\chi_{ij}G^{<}_{ji}(E) -H^\chi_{ji}G^{<}_{ij}(E) \right).
\label{IB}
\end{eqnarray}
The \textit{ring current} present in a specific ring can be calculated by summing up the currents in all bonds present in the ring along counter-clockwise direction. The MOM is calculated from the ring current by using the self-consistent Landauer-Keldysh formalism using Eq.\,\ref{Ms}, which also modifies the onsite energy for a given spin and thus is utilized to evaluate the total current at finite bias.

\section{Acknowledgement}
The authors would like to thank Nicolae Atodiresei for stimulating discussion. SG is co-funded by the European Union (Physics for Future - Grant Agreement No. 101081515). This work is supported by the Ministry of Education, Youth and Sports of the Czech Republic through the e-INFRA CZ (ID:90254).

\bibliography{Ref}

\clearpage
\pagebreak

\setcounter{page}{1}
\setcounter{figure}{0}
\renewcommand{\figurename}{Supplementary Figure}
\renewcommand{\tablename}{Supplementary Table}

\renewcommand{\thepage}{S\arabic{page}}
\renewcommand{\thefigure}{S\arabic{figure}}
\renewcommand{\thetable}{S\arabic{table}}
\renewcommand{\theequation}{S\arabic{equation}}

\section*{Supplementary Note 1: Structure of the isolated $\chi_+$-[7H] and $\chi_-$-[7H] molecules \label{A:Structure}}

\begin{figure}[h!]
\centering
\includegraphics[width=0.48\textwidth]{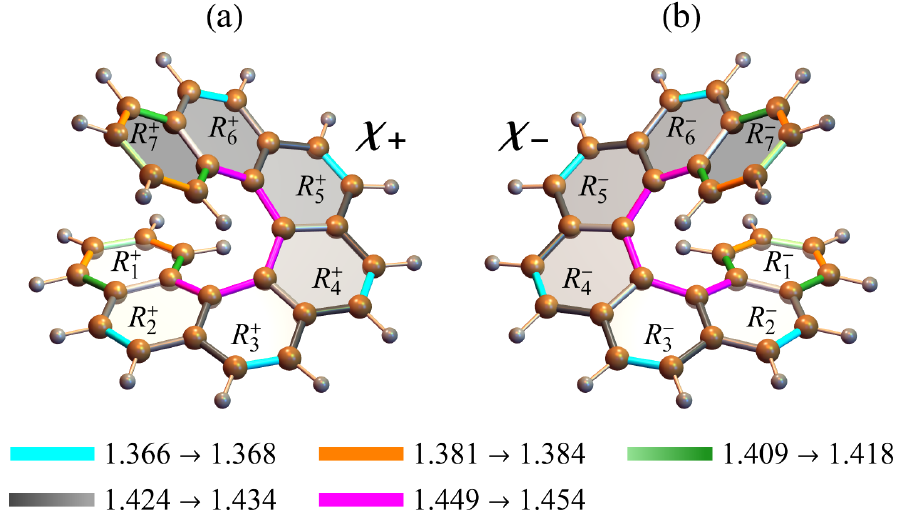}
\caption{Relaxed configuration of the heptahelicene molecule. (a) and (b) show the $\chi_+$-[7H] and $\chi_-$-[7H] molecule, where the large brown and small gray spheres represent the carbon and hydrogen atoms. The C-C bonds are colored according to their bond length. The color scale in Angstrom is shown at the bottom.}
\label{fig:Structure}                       
\end{figure}

Figure \ref{fig:Structure} shows the relaxed configuration of the isolated $\chi_\pm$-[7H] molecule obtained from the first-principle study (see Method section of the main text for computational details). From the color code, one can see that the longer bonds are always along the inner edge of the helix, while the shorter bonds are along the outer edge. For the first and last six-rings, the bond length variation does not follow the same pattern. When modeling these configurations with a simplified tight binding model, we will ignore the hydrogen atoms and focus on the $p_z$ orbitals of the C atoms. The strain caused by the bond length variation is modeled as a vector field which is assumed to be present in rings $R_2$ to $R_6$.

\begin{figure}[ht!]
\centering
\includegraphics[width=0.48\textwidth]{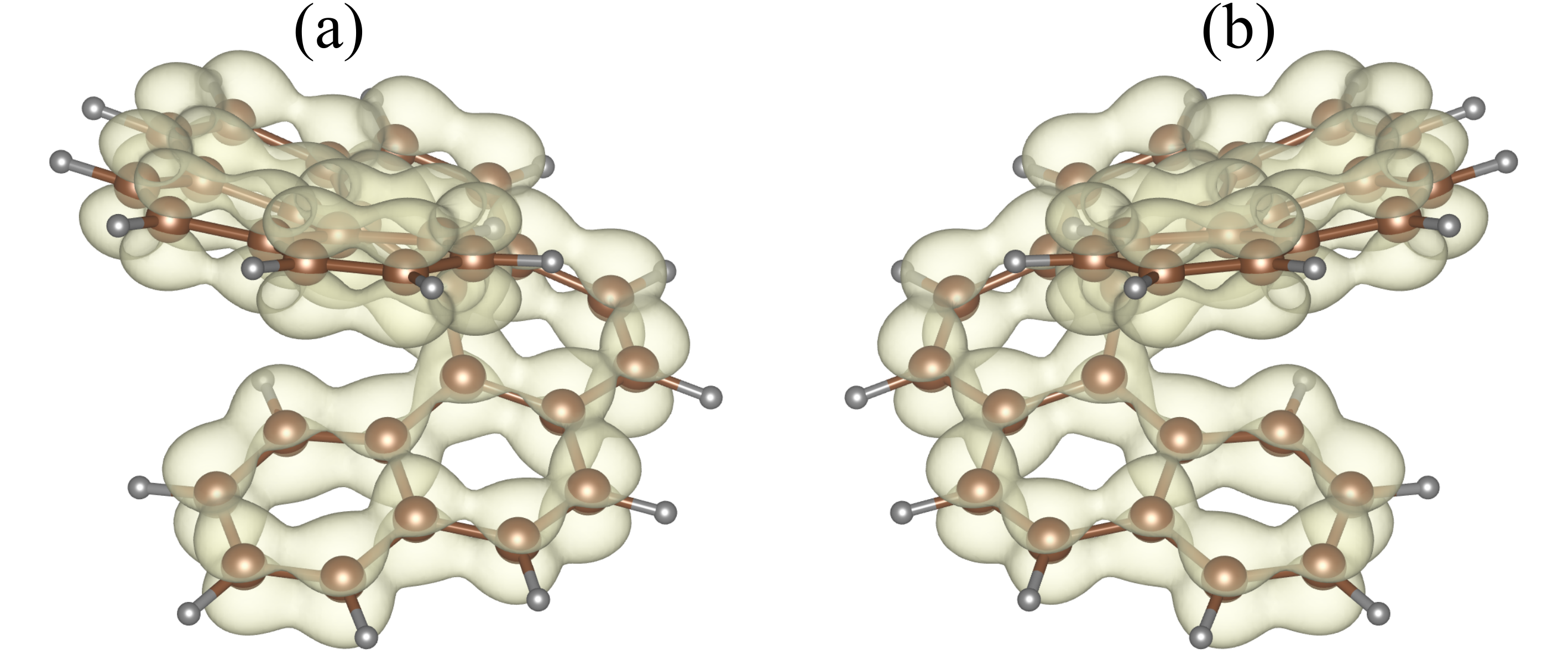}
\caption{Total density of states corresponding to all  states within the range of $E_{HO}$-2\,eV to $E_{LU}$+2\,eV where $E_{HO}$ and $E_{LU}$ are the highest occupied and lowest unoccupied energy eigenvalues. (a) and (b) show the $\chi_+$ and $\chi_-$ molecule. The isosurface represents a specific value of the density and is plotted with VESTA \cite{Momma2011}.}
\label{fig:LD}
\end{figure}

Our first-principle calculations show that the difference between the highest occupied and lowest unoccupied eigenvalues is 2.58\,eV, which is quite close to the value 2.23\,eV obtained from our simplified model excluding the H atoms. We further evaluate the integrated local density of states in the energy range from $E_{HO}-2$\,eV to $E_{LU}+2$\,eV, where $E_{HO}$ and $E_{LU}$ represent highest occupied and lowest unoccupied states, as these states play a major role in transport under bias (see Fig.\,\ref{fig:LD}). From Fig.\,\ref{fig:LD} one can see that this energy range is dominated by the C~$p_z$ orbitals. Therefore, to study the transport properties, one can neglect the H atoms and construct an effective Hamiltonian and consider only the C atoms, which is in line with our initial choice of the Hamiltonian based on C $p_z$ orbitals (see Eq.\,\ref{H} in the main text).

\section*{Supplementary Note 2: Molecule-electrode coupling \label{A:coupling}}

\begin{figure}[ht!]
\centering
\includegraphics[width=0.48\textwidth]{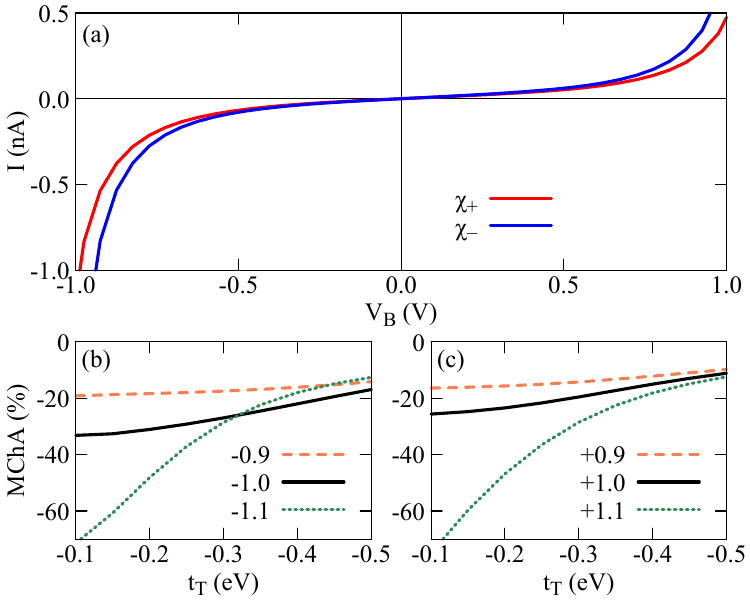}
\caption{$I$-$V$ curves for smaller molecule-electrode coupling and variation of the MChA as a function of the molecule-electrode coupling. (a) shows the $I$-$V$ curves for $t_B=-0.3$\,eV and $t_T=-0.1$\,eV. (b) and (c) show the variation of the MChA for different negative (b) and positive (c) bias voltages as a function of $t_T$, where the hopping parameter between the molecule and the bottom electrode was set to $t_B=3t_T$. Orange dashed, black, and green dotted lines show the MChA for bias voltage $\pm0.9$\,V, $\pm1.0$\,V, and $\pm1.1$\,V, respectively. For all curves the Fermi level was kept at $E_F=0.5$\,eV.}
\label{fig:TC}                       
\end{figure}

While the coupling parameter $t_B$ between the molecule and bottom electrode (substrate) depends on their chemical properties, the coupling parameter $t_T$ between the molecule and the top electrode (STM/AFM tip) strongly depends on their spatial separation as well. The current flowing through the molecule is quite sensitive to these coupling parameters especially in the tunnelling regime. In the main text, we consider comparatively large coupling parameters ($t_B=-1.5$\,eV, $t_T=-0.5$\,eV) to make all the effects clearly visible. Here we use smaller coupling parameters, which reproduce the experimentally observed current magnitudes (see Fig.\,\ref{fig:TC}a). For consistency we use $t_B=3t_T$ everywhere. Our results show that the current can decrease by three orders of magnitude when we decrease the coupling by a factor of 5. Such exponential decay is expected in the tunnelling regime when the transmission probability decays exponentially with the width of tunnelling barrier (separation between the molecule and the STM/AFM tip in this case) as well as with decreasing energy of the incident particle. For large bias voltage or near the resonant levels this dependence can be quite non-monotonic. Note that in absence of any scattering, the coupling to the electrodes is the only source of broadening of the resonant levels. Therefore, for reduced coupling with the electrodes, the resonant levels can be quite sharp and consequently the current can change significantly with the change of the applied bias voltage or the Fermi level. These features are clearly reflected in the variation of the MChA with respect to the $t_T$ for different applied bias voltages (Fig.\,\ref{fig:TC}b,c), which can also explain the large MChA values observed in experiments as well as their variations.

\pagebreak

\section*{Supplementary Note 3: Impact of bias polarity and chirality on the vector field-induced ring current \label{A:Ring}}

The aim of the following is to develop a simple physical picture showing how the vector field-induced ring current and MOM depend fundamentally on chirality and bias polarity, i.e., to provide a qualitative explanation of the key features discussed in the context of Figs.\,\ref{fig:IV} of the main text. To this end, we fix the Fermi level far from the resonant levels and consider small positive and negative bias voltages, such that the bias window does not approach the resonant levels. 

\begin{figure}[h!]
	\centering
	\includegraphics[width=0.40\textwidth]{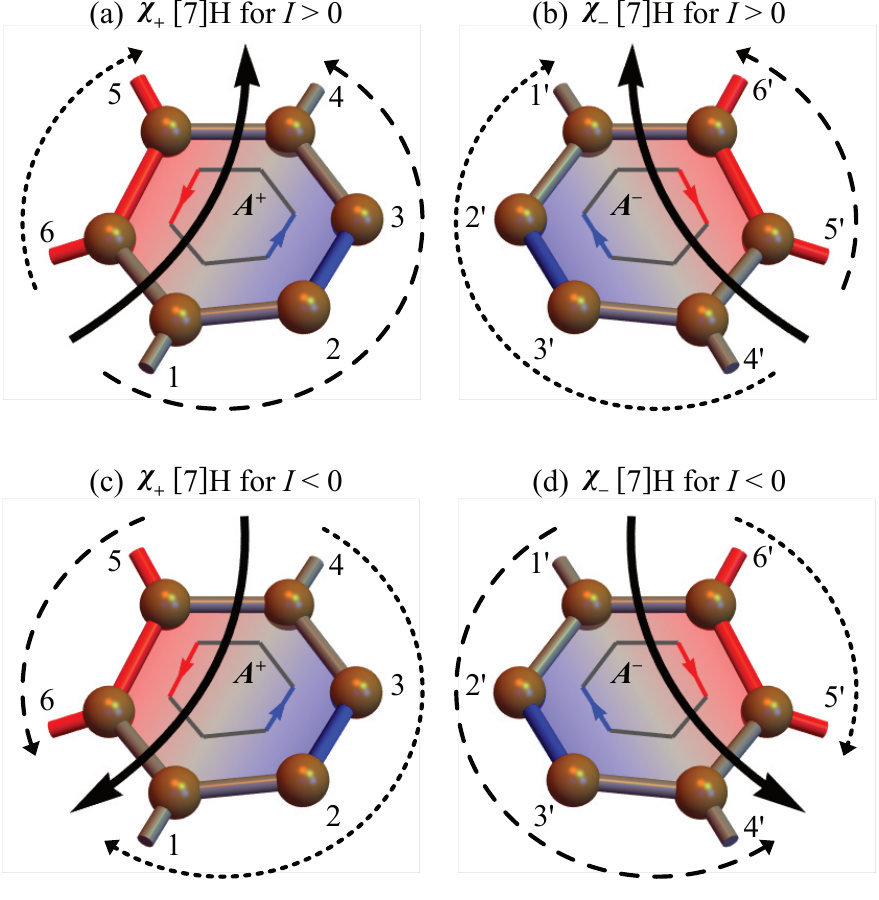}
	\caption{Vector field and current flow in the $4^{th}$ C six-ring of both [7]H enantiomers for positive (a,b) and negative (c,d) current flowing through the molecules. Solid black arrows indicate the direction of the corresponding electron motion.  
    Dashed and dotted arrows represent counter-clockwise (ccw) and clockwise (cw) current components, respectively. 
    Blue and red arrows indicate the direction of the vector field $\bm{A}^\chi$ along the outer and inner edge of the helix, similar to Figs.\,\ref{fig:h7}c,d of the main text.  
    The C atoms are numbered ccw from 1 to 6 and from 1' to 6', respectively.}
	\label{fig:A_Ring} 
\end{figure}

The gradient of the  bond length in a C six-ring gives rise to the vector field $\bm{A}^\chi$ shown within the rings in 
Fig.\,\ref{fig:h7}c,d of the main text. Figure \ref{fig:A_Ring} reproduces these configurations for both enantiomers as well as for positive (a,b) and negative (c,d) current through the molecule.
$\bm{A}^\chi$ is non-zero along the longer (red arrows) and shorter (blue arrows) bonds and zero for all other bonds (gray lines). 

The current flowing through the molecules splits in each C six-ring into a ccw (dashed arrows) and a cw (dotted arrows) component. The corresponding electronic wave functions propagating along the dashed and the dotted paths will acquire different phases. The phase difference $\Delta^\chi_I$ is calculated by subtracting the phase acquired along the cw path (dotted arrows) from that acquired along the ccw path (dashed arrows). The four configurations in Fig.\,\ref{fig:A_Ring} yield
\begin{widetext}
\begin{eqnarray}
	\nonumber
	\Delta^+_{I>0} =& \frac{e}{\hbar}\int_2^3\!\bm{A}^+\!\cdot\!d\bm{l}-\frac{e}{\hbar}\int_6^5\!\bm{A}^+\!\cdot\!d\bm{l} = \frac{e}{\hbar}\int_2^3\!\bm{A}^+\!\cdot\!d\bm{l}+\frac{e}{\hbar}\int_5^6\!\bm{A}^+\!\cdot\!d\bm{l} = \frac{e}{\hbar}\oint_+\!\bm{A}^+\!\cdot\!d\bm{l} &= +0.06\cdot 2\pi,\\ 
	\nonumber
	\Delta^-_{I>0} =&
	\frac{e}{\hbar}\int_{5'}^{6'}\!\bm{A}^-\!\cdot\!d\bm{l}-\frac{e}{\hbar}\int_{3'}^{2'}\!\bm{A}^-\!\cdot\!d\bm{l} = \frac{e}{\hbar}\int_{5'}^{6'}\!\bm{A}^-\!\cdot\!d\bm{l}+\frac{e}{\hbar}\int_{2'}^{3'}\!\bm{A}^-\!\cdot\!d\bm{l} = \frac{e}{\hbar}\oint_-\!\bm{A}^-\!\cdot\!d\bm{l} &= -0.06\cdot 2\pi,\\
	\nonumber
	\Delta^+_{I<0} =&
	\frac{e}{\hbar}\int_5^6\!\bm{A}^+\!\cdot\!d\bm{l}-\frac{e}{\hbar}\int_3^2\!\bm{A}^+\!\cdot\!d\bm{l} = \frac{e}{\hbar}\int_5^6\!\bm{A}^+\!\cdot\!d\bm{l}+\frac{e}{\hbar}\int_2^3\!\bm{A}^+\!\cdot\!d\bm{l} = \frac{e}{\hbar}\oint_+\!\bm{A}^+\!\cdot\!d\bm{l} &= +0.06\cdot 2\pi,\\
	\nonumber
	\Delta^-_{I<0} =&
	\frac{e}{\hbar}\int_{2'}^{3'}\!\bm{A}^-\!\cdot\!d\bm{l}-\frac{e}{\hbar}\int_{6'}^{5'}\!\bm{A}^-\!\cdot\!d\bm{l} = \frac{e}{\hbar}\int_{2'}^{3'}\!\bm{A}^-\!\cdot\!d\bm{l}+\frac{e}{\hbar}\int_{5'}^{6'}\!\bm{A}^-\!\cdot\!d\bm{l} = \frac{e}{\hbar}\oint_-\!\bm{A}^-\!\cdot\!d\bm{l} &= -0.06\cdot 2\pi.
\end{eqnarray}
\end{widetext}	

Note that reversing the bias polarity requires swapping all integration limits, which corresponds to a sign change for all phases acquired along the dotted and dashed paths. However, when calculating the {\it phase difference}, there is another sign change that cancels out the first one, since all paths are traversed in opposite directions (ccw $\leftrightarrow$ cw). In summery, we find that the phase difference is opposite for the two enantiomers and does not depend on the bias polarity. It can be written as $\Delta^\chi_{I{\genfrac {} {} {0pt} {2} {>} {<}0}}=\frac{e}{\hbar}\!\oint_\chi\!\bm{A}^\chi\!\cdot\!d\bm{l}\!=\!\chi0.06\cdot 2\pi$. 

The interference of the phase-shifted wave functions results in a lower current flow through the molecule compared to the situation without phase difference. The reduction depends on the magnitude of $\Delta^\chi_I$, but not on its sign. The magnitude of $\Delta^\chi_I$ also determines the strength of the vector field-induced ring current, while the sign indicates its direction (ccw or cw). As a consequence, the vector field-induced ring currents and the associated MOMs in the two enantiomers are opposite and independent of the bias polarity. These are the key ingredients that imprint  the molecule's handedness onto its transport properties and also lead to non-crossing $I$-$V$ curves.

\section*{Supplementary Note 4: Antisymmetric MOM and non-crossing \textit{I-V} curves  \label{A:A0}}

\begin{figure}[t]
\centering
\includegraphics[width=0.48\textwidth]{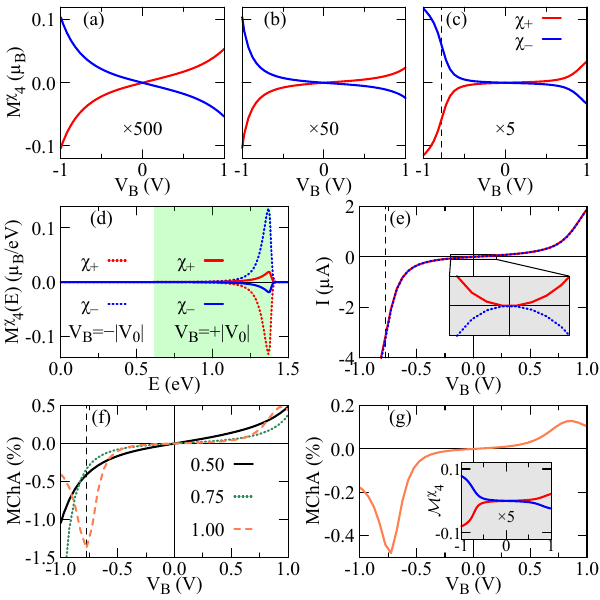}
\caption{MOM, MChA and \textit{I-V} curves for $\bm{A}^\chi=0$. (a-c) MOM of the $4^{th}$ six-ring for $E_F$ = (a) 0.5\,eV, (b) 0.75\,eV, and (c) 1.0\,eV, where red and blue correspond to the $\chi_+$ and $\chi_-$ enantiomers. The numbers below the curves denote the factor by which the actual value is multiplied. Vertical dashed line shows the voltage $V_0$ corresponding to maximum MChA (see Fig.\,\ref{fig:A0}f). 
(d) Distribution of the MOM for $\bm{A}^\chi=0$, $V_B=\pm V_0$, and $E_F=1.0$\,eV with the bias window shown in green.
(e) \textit{I-V} curves for $\bm{A}^\chi=0$. Inset shows the magnified non-crossing at $V_B=0$, where the red and blue lines show $I_{\chi}-\frac{I_+ + I_-}{2}$ for better visibility.
(f) MChA for $E_F = 0.5$\,eV (black), 0.75\,eV (green dotted) and 1.0\,eV (orange dashed), where the values are multiplied by a factor 100 (black), 10 (green dotted) and 1 (orange dashed). For $E_F=1.0$\,eV, the voltage $V_0$ corresponding to highest magnitude of MChA is marked by a vertical dashed line.
(g) Antisymmetric component of the MChA $\left(=\frac{\mathcal{I}_+ - \mathcal{I}_-}{\mathcal{I}_+ + \mathcal{I}_-} \right)$ and MOM ($\bm{\mathcal{M}}^\chi_4$) of the 4$^{th}$ six-ring calculated in presence of a finite $\bm{A}^\chi$ for $E_F=0.75$\,eV (see Figs.\,\ref{fig:CM}c,d in the main text for corresponding $\bm{M}^\chi_4$ and $\bm{\mathcal{M}}^\chi_4$).
}
\label{fig:A0}
\end{figure}

In the main text we show that there exists an antisymmetric component of the MOM (denoted as $\bm{\mathcal{M}}^\chi_4$). Here, we show that such a component can exist even without any structural distortion. To this end, we set $\bm{A}^\chi$=0 in Eq.\,\ref{H} of the main text and proceed in the same way as before to evaluate the MOM and \textit{I-V} characteristics (see Fig.\,\ref{fig:A0}). As expected, the MOM in this case is significantly smaller and antisymmetric with respect to the bias voltage (Figs.\,\ref{fig:A0}a-c), which stems from the absence of Fermi sea contribution (Fig.\,\ref{fig:A0}d). Since the MOM originates solely from the bias window, it follows the sign of the bias voltage and yields non-crossing \textit{I-V} curves (Fig.\,\ref{fig:A0}e). Consequently, the MChA also changes its sign with the bias voltage (Fig.\,\ref{fig:A0}f). 

To verify the equivalence between the MOM for $\bm{A}^\chi$=0 and the antisymmetric component of the MOM in presence of a finite $\bm{A}^\chi$, we compare the MOM of the $4^{th}$ ring for $E_F=1$\,eV and $\bm{A}^\chi=0$ (Fig.\,\ref{fig:A0}c) with $\bm{\mathcal{M}}^\chi_4$ calculated for finite $\bm{A}^\chi$ (inset of Fig.\,\ref{fig:A0}g). Note that the presence of $\bm{A}^\chi$ can also alter the position and width of the resonant levels. Therefore, for the calculation involving a finite $\bm{A}^\chi$, we choose $E_F=0.75$\,eV to compensate these changes. We also extract the antisymmetric component of the MChA in presence of $\bm{A}^\chi$ as $\frac{\mathcal{I}_+ - \mathcal{I}_-}{\mathcal{I}_+ + \mathcal{I}_-}$, where $\mathcal{I}_\chi=I_\chi(+\bm{A}^\chi)+I_\chi(-\bm{A}^\chi)$ (main graph of Fig.\,\ref{fig:A0}g). Our results show quite good qualitative agreement, which establishes the validity of our assumptions.

\pagebreak

\end{document}